\documentclass[aps,twocolumn,amsmath,amssymb,superscriptaddress]{revtex4-2}
\pdfoutput=1
\usepackage[colorlinks=true,linkcolor=blue,citecolor=blue,urlcolor=blue]{hyperref}

\usepackage{graphicx}
\usepackage{graphics}
\usepackage[export]{adjustbox}
\usepackage{bm}
\usepackage{amsfonts}
\usepackage{amssymb}
\usepackage{amsmath}
\usepackage{wasysym}
\usepackage{bbold}
\usepackage{stmaryrd}
\usepackage[usenames]{color}
\usepackage{colordvi}
\usepackage{units}
\usepackage{bbm}
\usepackage{booktabs}
\usepackage{array}
\usepackage{placeins}
\usepackage{epsfig}
\usepackage{changes}
\usepackage{braket}
\usepackage{tabularx}
\usepackage[normalem]{ulem}
\newcolumntype{L}[1]{>{\raggedright\arraybackslash}p{#1}} 
\newcolumntype{C}[1]{>{\centering\arraybackslash}p{#1}} 
\newcolumntype{R}[1]{>{\raggedleft\arraybackslash}p{#1}} 

\newcommand{\be}{\begin{equation}}
\newcommand{\ee}{\end{equation}}
\newcommand{\beqn}{\begin{eqnarray}}
\newcommand{\eeqn}{\end{eqnarray}}

\usepackage{orcidlink} 

\begin{document}

\title{Entanglement detection in postquench nonequilibrium states:\\
thermal Gibbs vs. generalized Gibbs ensemble}
\author{Ferenc Igl{\'o}i\,\orcidlink{0000-0001-6600-7713}}
\email{igloi.ferenc@wigner.hu}
\affiliation{Wigner Research Centre for Physics, Institute for Solid State Physics and Optics, H-1525 Budapest,  Hungary}
\affiliation{Institute of Theoretical Physics, University of Szeged, H-6720 Szeged, Hungary}
\author{Csaba Zolt\'an Kir\'aly}
\email{kiraly.csaba2000@gmail.com}
\affiliation{Institute of Theoretical Physics, University of Szeged, H-6720 Szeged, Hungary}
\date{\today}

\begin{abstract}
We use entanglement witnesses related to the entanglement negativity of the state to detect entanglement in the $XY$ chain in the  postquench states in the thermodynamic limit after a quench when the parameters of the Hamiltonian are changed suddenly. The entanglement negativity is related to correlations, which in the postquench stationary state are described by a generalized Gibbs ensemble, in the ideal case. If, however, integrability breaking perturbations are present, the system is expected to thermalize. Here we compare the nearest-neighbor entanglement in the two circumstances.

\end{abstract}

\pacs{}

\maketitle

\section{Introduction}
\label{sec:introduction}

Entanglement is a fundamental ingredient of quantum theory and has a central role in quantum information theory\cite{Horodecki2009Quantum,Guhne2009Entanglement,Friis2019}. For pure states this is directly related to correlations, while for mixed states entanglement has a more complex meaning. A quantum state is entangled, if its density matrix cannot be written as a mixture of product states. Deciding whether a state is entangled or not is a difficult problem. However, there are conditions that are necessary and sufficient for small systems., e.g. for $2\times 2$ (two-qubit) and $2\times 3$ bipartite systems \cite{Peres1996Separability,Horodecki1997Separability} and for multi-mode Gaussian states \cite{Giedke2001Entanglement}. There are also conditions that are sufficient conditions for entanglement for larger systems, but does not detect all entangled states.

Considering experiments usually only limited information about the quantum state is available and this is true for theoretical calculations for very large systems. Only those approaches for entanglement detection can be applied that require the measurement of a few observables. There are entanglement conditions that are linear in operator expectation values, these are the entanglement witnesses.They are operators that have a positive expectation value for all separable states. Thus, a negative expectation value signals the presence of entanglement. The theory of entanglement witnesses has recently been rapidly developing \cite{Horodecki1996Separability,Terhal2000Bell,Lewenstein2000Optimization,Acin2001Classification}. It is also known how to optimize a witness operator in order to detect the most entangled states \cite{Lewenstein2000Optimization}.

Apart from determining optimal entanglement witnesses, it is also important to find witnesses that are easy to measure in an experiment or possible to evaluate in a theoretical calculation.  From both point of views, witnesses based on spin chain Hamiltonians attracted considerable attention \cite{Toth2005EntanglementWitnesses,Toth2006Detection,Guhne2006Energy,Guhne2005Multipartite,Brukner2004MacroscopicB,Dowling2004Energy,Wu2005Entanglement}.  The energy-based witnesses have been used in various physical systems \cite{Vertesi2006Thermal,Sioli2012Towards,Siloi2013Quantum,Troiani2012Energy,Homayoun2019Energy,Troiani2013Detection}. However, it has been shown that the optimal witness for the thermal state of the chain is not necessarily the Hamiltonian \cite{Wu2005Entanglement}. Therefore it is recommended to consider another approach, based on {a family of witnesses that detect} entanglement whenever the entanglement negativity of the nearest-neighbor two-spin density matrix is nonzero \cite{Vidal2002Computable}, i.e., when the state violates the entanglement criterion based on the positivity of the partial transpose (PPT) \cite{Peres1996Separability,Horodecki1997Separability}.

For mixed states one generally considers thermal states and calculates a temperature bound, below which the state is entangled\cite{Toth2005EntanglementWitnesses,Toth2006Detection,Guhne2006Energy,Guhne2005Multipartite,Brukner2004MacroscopicB,Dowling2004Energy,Wu2005Entanglement}. Recently, however, the entanglement of other type of mixed states has also been considered, which are postquench nonequilibrium states. These states are obtained through such a protocol, when the quantum system is first placed to its ground state, then a quench is performed, when the parameters of the Hamiltonian change suddenly\cite{PhysRevA.2.1075,PhysRevA.3.2137,PhysRevLett.85.3233,PhysRevA.69.053616,RevModPhys.83.863}. Since the state is not an eigenstate of the new Hamiltonian, dynamics start. In the infinite time limit, the system approaches a stationary state, which is some mixture of the states appearing during the dynamics. If the Hamiltonian is non-integrable, the system is expected to be thermalized and the stationary state is described by a Gibbs ensemble with an effective temperature  \cite{Sotiriadis_2012,PhysRevA.79.021608,Sotiriadis_2009,PhysRevA.78.013626,PhysRevLett.100.100601,PhysRevLett.101.063001,PhysRevLett.100.030602,PhysRevLett.98.210405,PhysRevLett.97.156403,PhysRevLett.96.136801,PhysRevLett.98.050405}, see however Refs.~\cite{PhysRevE.93.032116,Larson_2013,PhysRevLett.106.025303,Olshanii2012}. For integrable systems, such as the transverse Ising chain, $XY$ and $XXY$ chains, the stationary state is assumed to be described by a so-called Generalized Gibbs Ensemble (GGE)  \cite{Vidmar_2016,Ilievski_2016,PhysRevLett.115.157201,PhysRevA.91.051602,Pozsgay_2014a,Pozsgay_2014,PhysRevA.90.043625,PhysRevLett.113.117203,PhysRevLett.113.117202}, for which different effective temperatures are assigned to each conserved quantities. This type of description has been exactly calculated for the quantum Ising chain \cite{Calabrese_2012}, and a similar formalism is conjectured for the $XY$ chain \cite{Blass_2012}. 

In experiments one cannot realize such systems, which are purely integrable, since weak integrable breaking perturbations are always unavoidable. In the presence of a perturbation that breaks integrability, usual thermalization is again expected to take
place. However, if the perturbation is small, the process
may require a long time. On a finite time scale, the dynamics is approximately described by the evolution under the integrable unperturbed Hamiltonian. The system
initially relaxes to a stationary state of the unperturbed
Hamiltonian, which is called prethermalization, while genuine thermalization only occurs at later times \cite{PhysRevLett.93.142002,PhysRevB.84.054304,Langen_2016}. This later
thermalization is typically involes a thermalization time $\tau \sim \lambda^{-2}$, where $\lambda$ is the perturbation strength \cite{PhysRevX.9.021027,PhysRevB.104.184302}. However, for specific Hamiltonians this timescale can be much longer\cite{Abanin2017,surace2023weak}. 

In this paper, we aim to compare the entanglement properties of a prethermalized and a genuine thermalized state. For this purpose we use the $XY$ model, which is exactly solvable and consider some type of integrable breaking perturbation. For the unperturbed model several entanglement based properties have been studied recently \cite{Osterloh2002,PhysRevA.66.032110,Patan__2007,PhysRevB.89.134101,PhysRevA.88.052305} and also the postquench nonequilibrium stationary state is analysed by energy-based and negativity-based entanglement witnesses\cite{PhysRevResearch.5.013158}.  Here we let switch on an integrable breaking perturbation and repeat the calculation.

Our paper is organized as follows. In Sec.~\ref{sec:model}, we introduce the $XY$ model, present its free-fermion representation, calculate thermal averages and present its conjectured GGE after a global quench. In Sec.~\ref{sec:witness}, the negativity-based entanglement witness is described. In Sec.~\ref{sec:post_quench}, the bounds for postquench states are calculated and the entangled areas are compared for prethermalized and genuine thermalized states. In Secs.~\ref{sec:discussion} we close our paper with a discussion.


\section{Model and methods}
\label{sec:model}

Here we consider the $XY$ spin-chain defined by the Hamiltonian
\begin{align}
\displaystyle{\cal H}_{XY}=&-\sum_{l=1}^L \left[\dfrac{1+\gamma}{2} \sigma^x_l \sigma^x_{l+1}+\dfrac{1-\gamma}{2} \sigma^y_l \sigma^y_{l+1}\right]\nonumber\\
&-h \sum_{l=1}^L \sigma^z_l+\lambda \displaystyle{\cal V}(\{\pmb{\sigma}\})\;,
\label{hamilton}
\end{align}
in terms of the $\sigma_l^{x},$ $\sigma_l^{y},$ and $\sigma_l^{z}$  Pauli spin operators acting on the spin at site $l,$ and  $\sigma_{L+1}^{\alpha}\equiv\sigma_{1}^{\alpha}$ for $\alpha=x,y,z.$ We mention that the special case $\gamma=1$ represents the transverse Ising model, and for $h=0$, $\gamma=0$ the Hamiltonian reduces to the $XX$ chain.

Later we extend the Hamiltonian with a general integrability breaking term:
\be
\displaystyle{\cal H}=\displaystyle{\cal H}_{XY}+\lambda \displaystyle{\cal V}(\{\pmb{\sigma}\})\;,
\label{hamiltonV}
\ee
This term, $\displaystyle{\cal V}(\{\pmb{\sigma}\})$ can c.f. contain interaction between more distant neighbours, but we do not specify its form, the r\v{o}le of this perturbation for $\lambda \ll 1$ is to ensure thermalization of the model after a quench for sufficiently long time, $\sim \lambda^{-2}$.

\subsection{Integrable model - $\lambda=0$}
\label{sec:integrable}

In detail we consider the integrable model with $\lambda=0$ and using standard techniques \cite{Lieb1961Two,Pfeuty1970The} it is expressed in terms of fermion creation and annihilation operators $\eta^{\dag}_p$ and $\eta_p$ as
\be
{\cal H}_{XY}=\sum_p \varepsilon\left(p\right)\left(\eta^{\dag}_p \eta_p-\frac{1}{2}\right),
\ee
where the sum runs over $L$ quasi-momenta,  which are equidistant in $[-\pi,\pi]$ for periodic  boundary conditions . The energy of the modes is given by \cite{PhysRevA.3.2137,PhysRevA.2.1075,Blass_2012}
\be
\varepsilon\left(p\right)=2\sqrt{\gamma^2 \sin^2 p+\left(h-\cos p\right)^2}
\label{eps}
\ee
and the Bogoliubov angle $\varTheta_p$ diagonalizing the Hamiltonian is given by 
\be\tan\varTheta_p=-\gamma\sin p/\left(h-\cos
p\right).\ee
The ground state is the fermionic vacuum, its energy being
\be
E_0=-\sum_p \frac{\varepsilon\left(p\right)}{2}\;.
\ee
The model has a so called \textit{disorder line} at
\be
h_d(\gamma)=\sqrt{1-\gamma^2}.\label{eq:hd}
\ee
along which the ground state of the model is a product state, i.e. not entangled.
For $h<h_d(\gamma)$  ($h>h_d(\gamma)$) the long-range two-point correlation functions have an (don't have) oscillatory behavior, while at $h=h_d(\gamma)$ they are constant \cite{PhysRevA.2.1075,PhysRevA.3.2137}.

At finite temperature, $T>0$ the average value of the energy is given by
\be
\langle {\cal H} \rangle_T=-\sum_p t(p,T)\frac{\varepsilon\left(p\right)}{2},
\label{E_T}
\ee
with 
\be
t(p,T)=\tanh\left(\frac{\varepsilon\left(p\right)}{2T}\right).
\label{t(p)}
\ee
(Here and in the following we set $k_B=1$.)

In the thermodynamic limit, $L \to \infty$, the two-point correlation functions are calculated in Refs.~\cite{PhysRevA.3.2137,PhysRevA.2.1075} and the nearest-neighbor correlations are given by:
\begin{align}
\langle \sigma_l^x \sigma_{l+1}^x \rangle_T&=g_c-g_s,\nonumber\\
\langle \sigma_l^y \sigma_{l+1}^y \rangle_T&=g_c+g_s,\nonumber\\
\langle \sigma_l^z \sigma_{l+1}^z \rangle_T&=g_0^2-g_c^2+g_s^2,
\label{nn_corr}
\end{align}
with
\begin{align}
g_c&=\frac{1}{\pi}\int_{-\pi}^{\pi} {\rm d} p \cos p (\cos p -h)~t(p,T)\varepsilon^{-1}\left(p\right),\nonumber\\
g_s&=-\gamma \frac{1}{\pi}\int_{-\pi}^{\pi} {\rm d} p \sin^2 p ~t(p,T)\varepsilon^{-1}\left(p\right),\nonumber\\
g_0&=\frac{1}{\pi}\int_{-\pi}^{\pi} {\rm d} p (h-\cos p) ~t(p,T)\varepsilon^{-1}\left(p\right).
\label{par_corr}
\end{align}

\section{Nonequilibrium stationary states after a quench}
\label{sec:quench}

We consider global quenches at zero temperature, which suddenly change the parameters of the Hamiltonian from $\gamma_0$, $h_0$ for $t<0$ to $\gamma$, $h$ for $t>0$, keeping however the value of. For $t<0$ the system is assumed to be in equilibrium, i.e., in the ground state $\left|\Phi_0\right\rangle$ of the Hamiltonian ${\cal H}_0$ with parameters $\gamma_0$ and $h_0.$ After the quench, for $t>0$, the state evolves coherently according to the new Hamiltonian ${\cal H}$ as
\be
\left|\Phi_0(t)\right\rangle=\exp(-i{\cal H}t)\left|\Phi_0\right\rangle.
\label{Phi(t)}  
\ee
Correspondingly, the time evolution of an operator in the Heisenberg picture is
\be
\sigma_l\left(t\right)=e^{i{\cal H}t} \sigma_l e^{-i{\cal H}t}.
\ee

After large enough time and in the thermodynamic limit the system is expected to reach a stationary state, 
\be
{\bm\rho}_{q}=\lim_{\tau\rightarrow\infty}\frac1 \tau \int_{0}^{\tau} e^{-i {\cal H}t} |\Phi_0\rangle\langle \Phi_0| e^{+i{\cal H}t}  {\rm d}t,
\ee
So that, for an observable ${\cal O}$ the stationary value is given by 
\be
\langle {\cal O} \rangle_{\rm st}={\rm Tr}({\bm\rho}_{q} {\cal O})\;.
\ee

\subsection{Stationary values in the integrable model}
\label{sec:stat_integrable}

In the integrable model with $\lambda=0$ the energy of the system after the quench is given as
\be
\langle \Phi_0| \displaystyle{\cal H} |\Phi_0 \rangle=\sum_p \varepsilon\left(p\right)\left(\langle \Phi_0|\eta^{\dag}_p \eta_p|\Phi_0 \rangle-\frac{1}{2}\right),
\label{E_Phi}
\ee
where  the occupation
probability of mode $p$ in the initial state $\left|\Phi_0\right\rangle$ is given as 
\be
f_p=\left\langle \Phi_0\right|\eta^{\dag}_p \eta_p \left|\Phi_0\right\rangle.
\ee 
For the $XY$ model it is expressed through the difference
$\Delta_p=\varTheta^0_p-\varTheta_p$ of the Bogoliubov angles as
\be 
f_p=\tfrac{1}{2}\left(1-\cos \Delta_p\right)
\label{f_p}
\ee 
with the cosine of the difference $\Delta_p$ given as 
\be
\scalebox{0.925}
{
$\cos \Delta_p=4\dfrac{\left(\cos p -h_0\right)\left(\cos p -h\right)+\gamma\gamma_0\sin^2 p}{\varepsilon\left(p\right)\varepsilon_0\left(p\right)},$
}
\label{Delta}
\ee
where the index $0$ refers to quantities before the quench \cite{Blass_2012}. In the thermodynamic limit, Eq.~(\ref{E_Phi}) can be rewritten as
\be
\frac{\langle \Phi_0| \displaystyle{\cal H} |\Phi_0 \rangle}{L}=-\frac{1}{4\pi}\int_{-\pi}^\pi \varepsilon(p)\cos \Delta_p {\rm d}p.\label{eq:quenchth}
\ee
The fermions with occupation probability $f_p$ are quasiparticles, which are created homogeneously in space and the corresponding wave-packets move ballistically with constant velocity. Such a wave packet is well described by a sharp kink excitation, if the quench is performed deep into the ordered phase. For quenches close to the critical point the kinks are not sharply localized and the domain walls have a finite extent of the order of the equilibrium correlation length. In the thermodynamic limit this effect can be taken into account by using an effective occupation probability:
\be
f_p \to {\tilde f}_p=-\frac{1}{2}\ln|\cos \Delta_p|\;,
\ee
so that in leading order ${\tilde f}_p=f_p + {\cal O}(f_p^2)$.

 In the stationary  state, due to conserved symmetries, averages of correlations are described by a Generalized Gibbs Ensemble (GGE)\cite{Vidmar_2016,Ilievski_2016,PhysRevLett.115.157201,PhysRevA.91.051602,Pozsgay_2014,PhysRevA.90.043625,PhysRevLett.113.117203,PhysRevLett.113.117202}. In this case to each fermionic mode an effective temperature, $T_{\rm eff}(p)$ is attributed through the relation \cite{Blass_2012}
\be
\tanh\left(\frac{\varepsilon\left(p\right)}{2T_{\rm eff}(p)}\right)=e^{-2{\tilde f}_p}=|2f_p-1|=|\cos \Delta_p|.
\label{T_eff}
\ee
In this way the nearest-neighbor correlations in the stationary state can be obtained as in section \ref{sec:integrable}, just replacing $t(p,T)$ defined in Eq.~(\ref{t(p)}) by $|\cos \Delta_p|$
\be
t(p,T) \to |\cos \Delta_p|.
\label{t_to_cos}
\ee
In particular, we have to apply Eq.~(\ref{t_to_cos}) for the correlation functions in Eqs.~(\ref{nn_corr}) and (\ref{par_corr}).

\subsection{Stationary values in the non-integrable model}
\label{sec:stat_nonintegrable}

In the non-integrable model with $0<\lambda \ll 1$ on finite time scale prethermalization takes place and the quasi-stationary state is described by a GGE, as explained in the previous subsection \ref{sec:stat_integrable}. After sufficiently long time, however, the system is expected to be thermalized and the genuine stationary state is expected to be described by a Gibbs ensemble. Possible ways to define a  characteristic thermalization temperature, $T_{th}$, have been discussed in several papers\cite{Sotiriadis_2009,PhysRevLett.102.127204,PhysRevB.82.144302,PhysRevA.80.053607,PhysRevB.83.094431,PhysRevB.84.212404}. 
%
%
%
%
After the quench at the prethermalization state the different fermionic modes in the system are characterised by a set of effective temperatures in Eq.(\ref{T_eff}) and at later times in the thermalised stationary state some average of these effective temperatures is expected. For small $\lambda$ the average value of the energy remains the same, from which the following condition for the thermalization temperature, $T_{th}$ follows:
%
\be
\int_{-\pi}^\pi \varepsilon(p)\tanh\left(\frac{\varepsilon\left(p\right)}{2T_{\rm eff}(p)}\right) {\rm d}p=\int_{-\pi}^\pi \varepsilon(p) \tanh\left(\frac{\varepsilon\left(p\right)}{2T_{th}}\right){\rm d}p\;.
\label{Tth2}
\ee
%

\section{Negativity-based entanglement witness}
\label{sec:witness}

Generally an operator ${\cal W}$ is called an entanglement witness, if its expectation value, $\langle {\cal W} \rangle$ satisfies the following requirements \cite{Lewenstein2001Characterization,Terhal2002Detecting}.

(i)  $\langle {\cal W} \rangle \ge 0$  for all separable states,

(ii) $\langle {\cal W} \rangle < 0$ for some entangled state.
 
Such a state is detected by the witness as entangled. Entanglement witnesses have been used in various physical systems to verify the presence of entanglement \cite{Bourennane2004Experimental, Walther2005ExperimentalOneWay,Kiesel2005Experimental,Wieczorek2009Experimental,Prevedel2009Experimental,Gao2010Experimental,Genuine2019Gong,Haffner2005Scalable,14qubit2011Monz,Feldman2008,Garttner2018Relating}. We mention that a single entanglement witness cannot detect all entangled states. 

Here we consider one of the most important entanglement witness, which is connected to the partial transpose of the density matrix\cite{Peres1996Separability,Horodecki1997Separability} and to the entanglement negativity\cite{Vidal2002Computable}. For a bipartite density matrix given as
\be
{\bm \rho}=\sum_{kl,mn}{\bm \rho}_{kl,mn} \ket{k}\bra{l}\otimes \ket{m}\bra{n}
\ee
the partial transpose according to first subsystem is defined by exchanging subscripts $k$ and $l$ as 
\be
{\bm \rho}^{T_A}=\sum_{kl,mn}{\bm \rho}_{lk,mn} \ket{k}\bra{l}\otimes \ket{m}\bra{n}.
\ee
It has been shown that for separable quantum states  \cite{Peres1996Separability,Horodecki1996Separability}
\be
{\bm \rho}^{T_A}\ge 0
\ee
holds. Thus, if ${\bm \rho}^{T_A}$ has a negative eigenvalue then the quantum state is entangled. For $2\times2$ and $2\times3$ systems, the PPT condition detects all entangled states \cite{Horodecki1996Separability}. For systems of size $3\times3$ and larger, there are PPT entangled states \cite{Horodecki1997Separability,Horodecki1998Mixed-State}. The entanglement negativity \cite{Vidal2002Computable} is defined as 
\be
{\cal N}(\bm{\rho})=2{\rm max}(0,-{\rm min}(\mu_{\nu})),
\label{negativity}
\ee
where $\mu_{\nu}$ are the eigenvalues of the partial transpose $\bm{\rho}^{T_A}.$

Let us turn to $XY$ chains and consider the nearest-neighbor reduced density matrix, $\bm{\rho}$, which is defined in the $\sigma^z$ basis. As described in details in Ref.\cite{Wu2005Entanglement,PhysRevResearch.5.013158} due to symmetries of the problem $\bm{\rho}$ is a direct sum of two $2 \times 2$ matrices and the same property holds also for the partial tranpose ${\bm \rho}^{T_A}$. The minimal eigenvalues of the $2 \times 2$ submatrices can be calculated by second order quadrature in terms of the matrix-elements of $\bm{\rho}$. The later for the XY and Heisenberg spin chains  can be expressed through nearest-neighbor correlations~{\cite{PhysRevA.66.032110}. The final results for the minimal eigenvalues are given by\cite{Wu2005Entanglement,PhysRevResearch.5.013158}:
\begin{align}
&\mu_{\rm min}^{(1)}=\frac{\langle \sigma_l^z \sigma_{l+1}^z \rangle+1}{4}\nonumber\\
&-\frac{1}{4}\sqrt{(\langle \sigma_l^z\rangle+\langle \sigma_{l+1}^z \rangle)^2+(\langle \sigma_l^x \sigma_{l+1}^x \rangle+\langle \sigma_l^y \sigma_{l+1}^y \rangle)^2}\;,\label{eq:mumin1_corr}
\end{align}}
and
\be
\mu_{\rm min}^{(2)}=-\frac{1}{4}(\langle \sigma_l^x \sigma_{l+1}^x \rangle-\langle \sigma_l^y \sigma_{l+1}^y \rangle +\langle \sigma_l^z \sigma_{l+1}^z \rangle-1).
\label{witness}
\ee
The entanglement witness related to $\mu_{\rm min}^{(2)}$ is given by:
\be
{\cal W}_N=-\frac{1}{4}( \sigma_l^x \sigma_{l+1}^x -\sigma_l^y \sigma_{l+1}^y  + \sigma_l^z\sigma_{l+1}^z -\openone),
\label{neg_witness1}
\ee
whereas the same for $\mu_{\rm min}^{(2)}$ is more complicated and derived in\cite{PhysRevResearch.5.013158}.

In order to proceed, we need to know that the partial transposition of a two-qubit state has at most one negative eigenvalue and all the eigenvalues lie in $[-1/2,1]$\cite{PhysRevA.58.826,PhysRevA.87.054301}. Thus, only one of the $\mu_{\rm min}^{(1)}$ and $\mu_{\rm min}^{(2)}$ can be negative, and when they are equal to each other, they must be non-negative and the state must be separable.

\section{Entanglement in nonequilibrium postquench states}
\label{sec:post_quench}

In this section, we consider global quenches in the system, as described in Sec.~\ref{sec:quench} and study the entanglement properties of nonequilibrium stationary states, which are obtained in the large-time limit after the quench. First we consider the prethermalized state, which is obtained by setting formally $\lambda=0$ in Eq.(\ref{hamiltonV}) and having the integrable model. In this part to calculate averages we use the GGE protocol and assign different effective temperatures to each fermionic modes, as described in Eqs.~(\ref{T_eff}) and (\ref{t_to_cos}). Afterwards we let to switch on a small perturbation, $\lambda \ll 1$, and consider the genuine thermalized state in which there is a unique thermalization temperature as defined in Eq.(\ref{Tth2}).

We have calculated of postquench states detected as entangled by the entanglement negativity-based witness using the two minimal eigenvalues, $\mu_{\rm min}^{(1)}$ and $\mu_{\rm min}^{(2)}$ in Eqs(\ref{eq:mumin1_corr}) and (\ref{witness}), respectively. These are presented {in Fig.~\ref{fig_1} having the entangled regions where $\mu_{\rm min}^{(2)}<0$, and in Fig.~\ref{fig_2}  where in the entangled regions  $\mu_{\rm min}^{(1)}<0$. Entanglement detected areas in the prethermalized  states are coloured by violet and in the genuine thermalized states by green. Overlapping areas are striped and contain both colours.
\begin{figure*}[t]
\centering
  \begin{tabular}{@{}cc@{}}
\includegraphics[width=1.\columnwidth,angle=0]{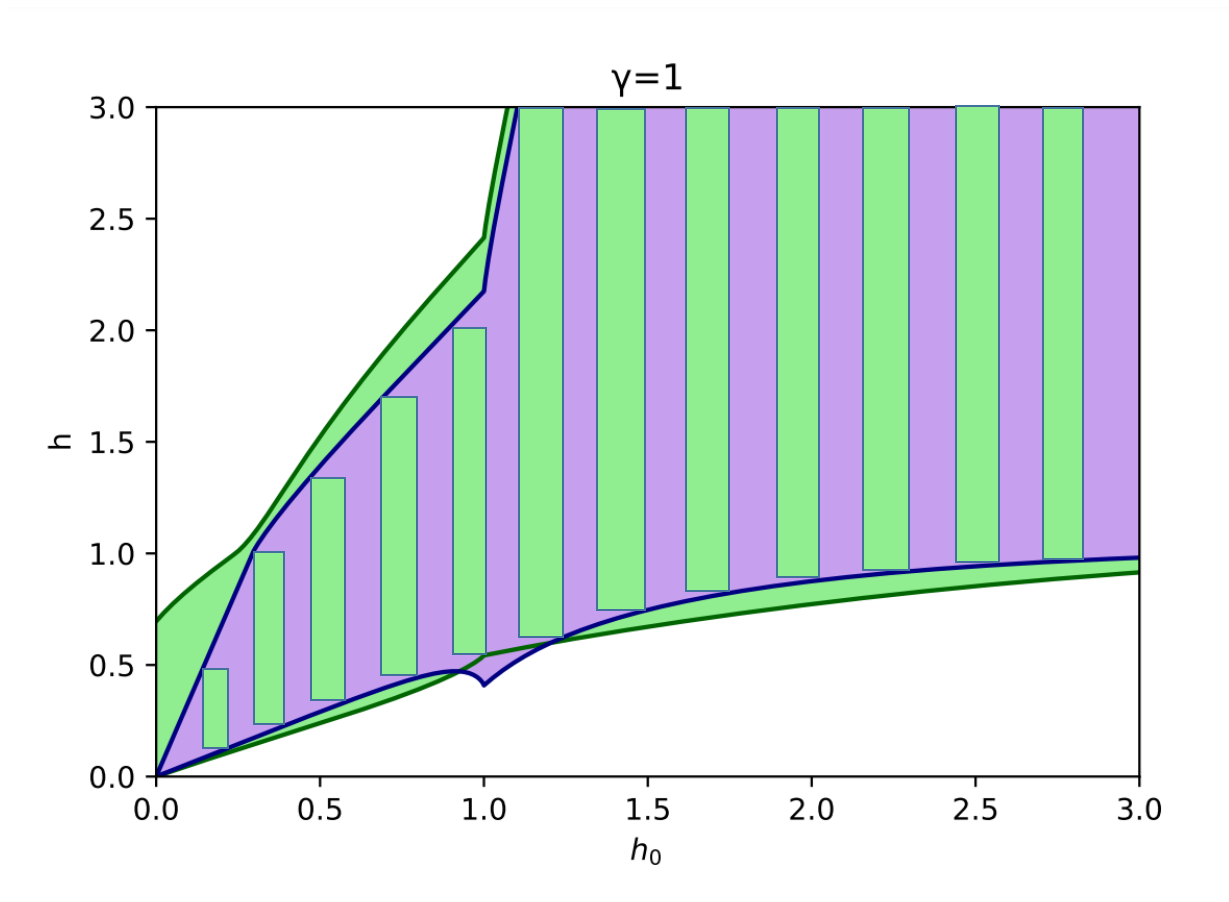} &
\includegraphics[width=1.\columnwidth,angle=0]{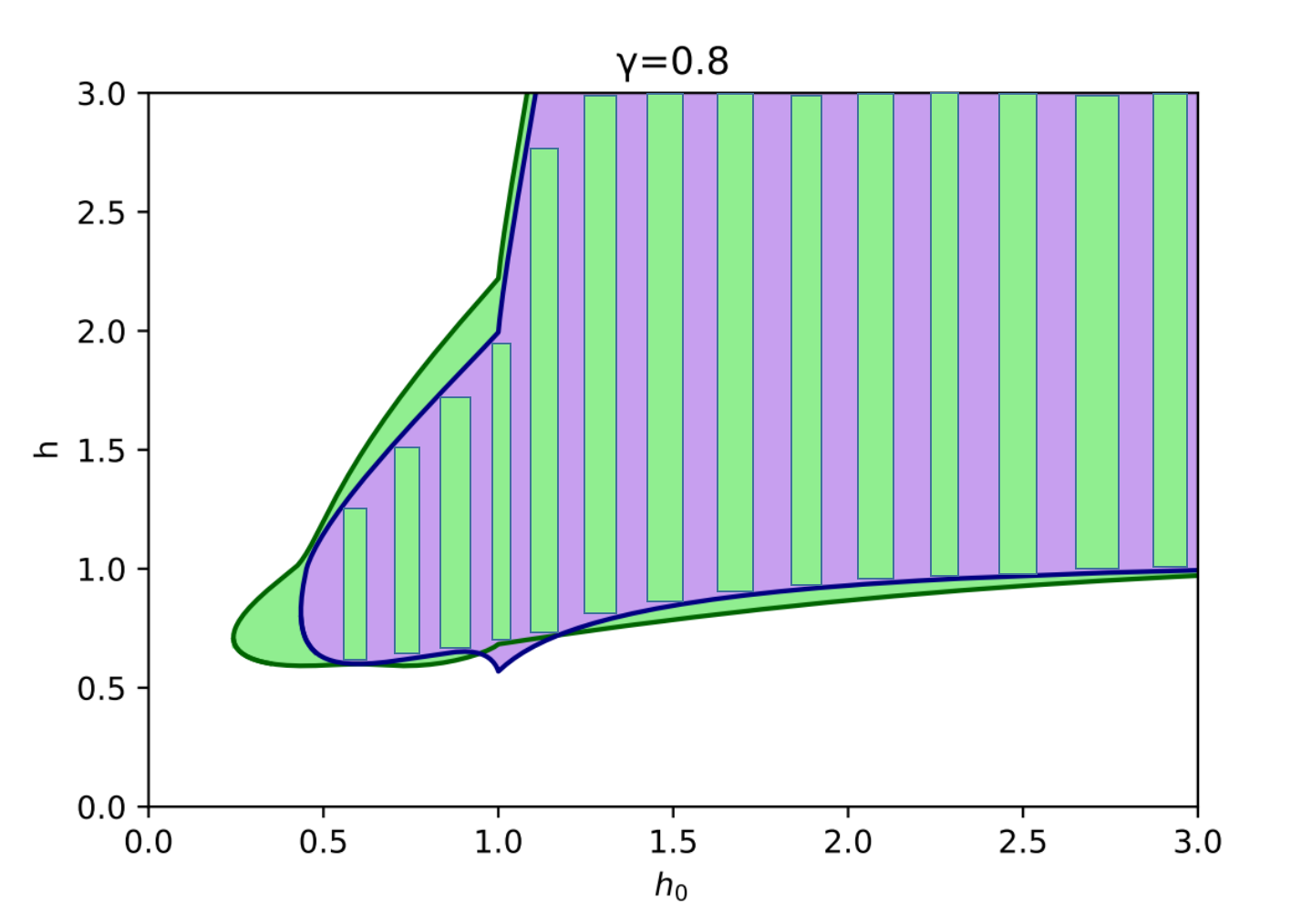} \\
\hspace{0.7cm}(a) & \hspace{1cm}(b)\\
&\\
\includegraphics[width=1.\columnwidth,angle=0]{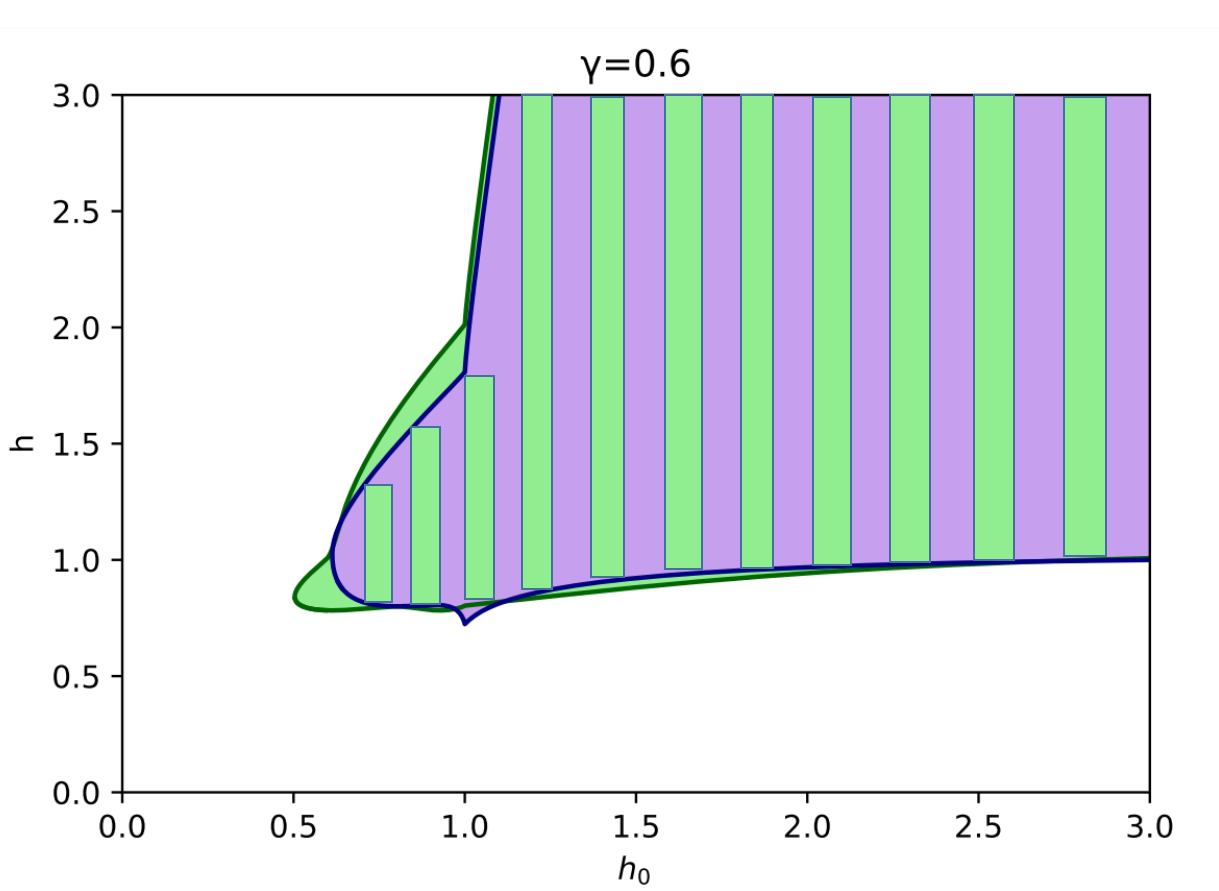} &
\includegraphics[width=1.\columnwidth,angle=0]{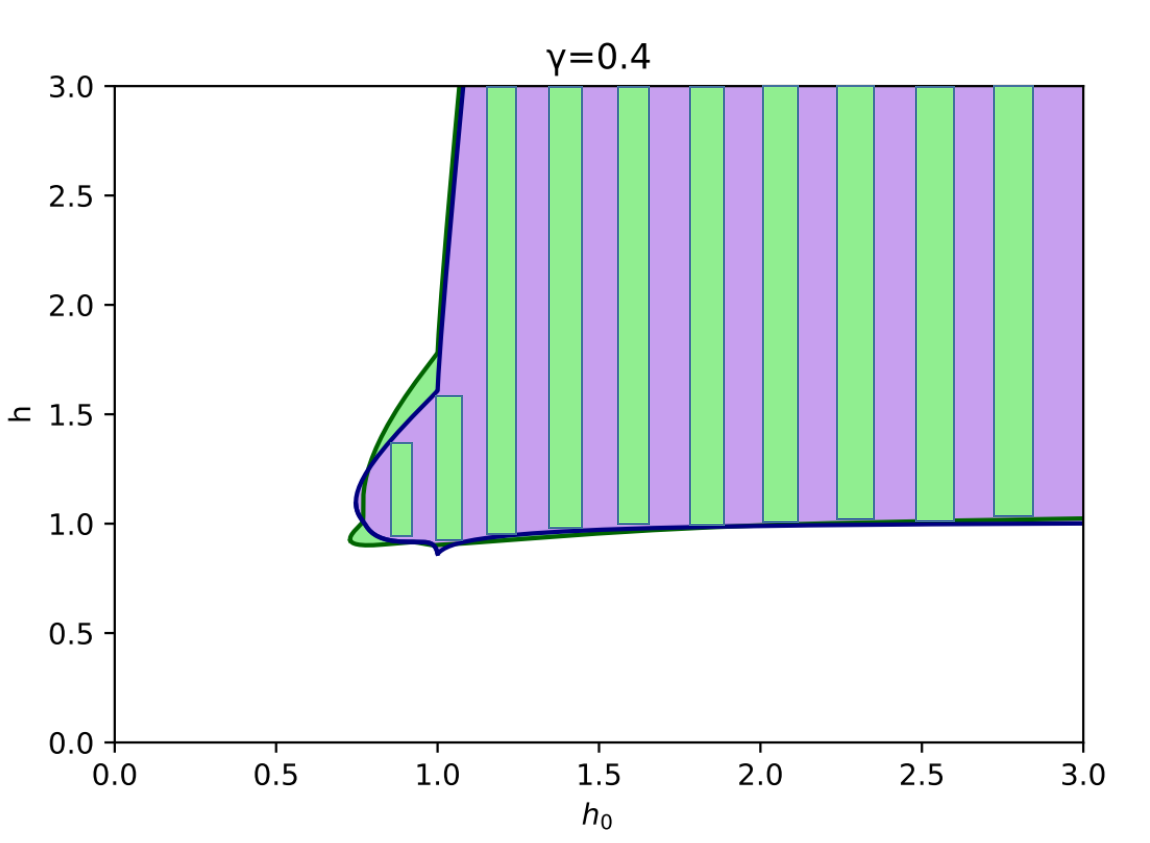} \\
\hspace{0.7cm}(c) & \hspace{1cm}(d)\\
&\\
\includegraphics[width=1.\columnwidth,angle=0]{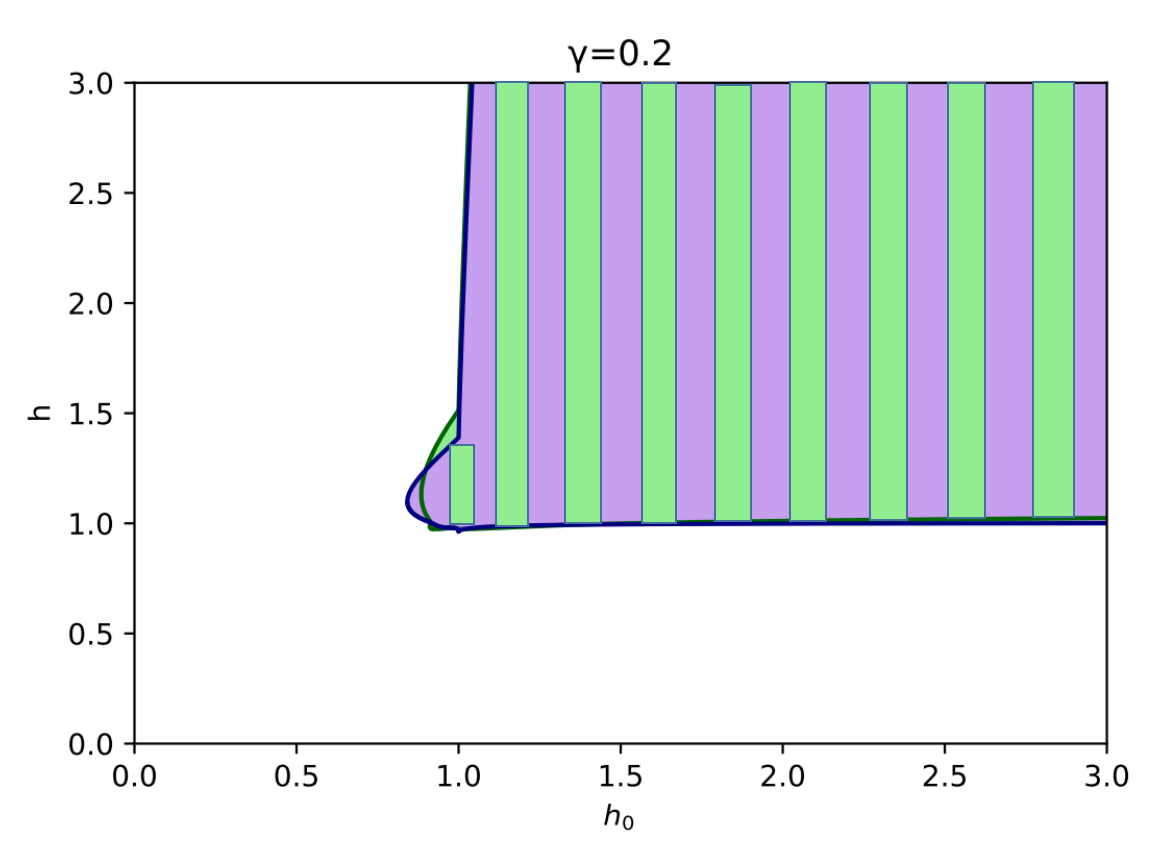} &
\includegraphics[width=1.\columnwidth,angle=0]{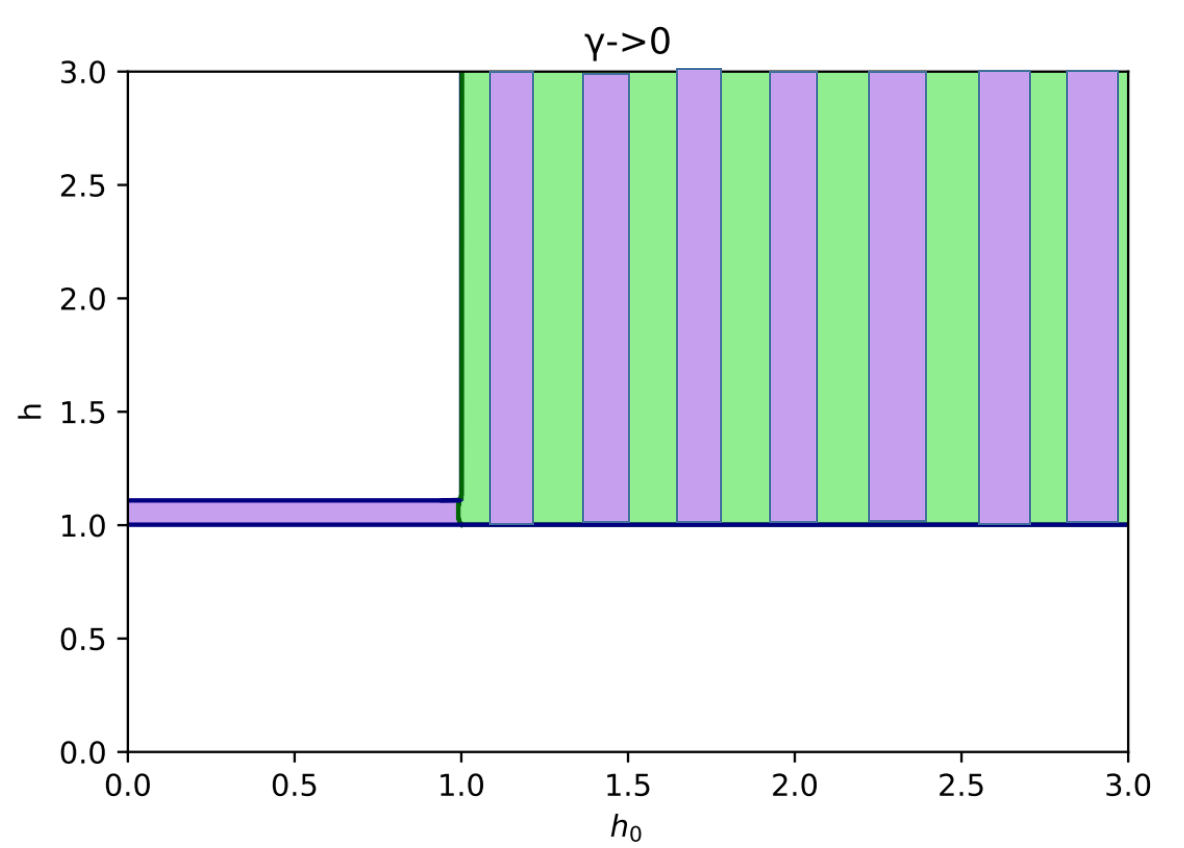} \\
\hspace{0.7cm}(e) & \hspace{1cm}(f)
  \end{tabular}
\caption{Postquench states after a sudden quench protocol $(h_0,\gamma) \to (h,\gamma)$ in the $XY$ chain.  { Entanglement is detected in the postquench state by the negativity-based method using $\mu^{(1)}_{\rm min}$ in Eq.~(\ref{eq:mumin1_corr}): in the prethermalized state (violet) and in the genuine thermalized state (green). Overlapping areas are striped.}
\label{fig_1}}	
\end{figure*} 

In Fig.~\ref{fig_1}(a), we consider the case with $\gamma=1$ and hence we have a quantum Ising chain. In this case, the negativity-based witness with {$\mu^{(2)}_{\rm min}$ is applicable in the whole phase diagram, consequently in Fig.~\ref{fig_2} (a) there is no entangled region detected.  In Fig.~\ref{fig_1}(b), (c), (d),(e) and (f),  we have $\gamma<1,$ and the condition $\mu_{\rm min}^{(1)}<\mu_{\rm min}^{(2)}$ is fulfilled in a part of the phase diagram. Thus, entangled postquench states are also detected based on $\mu^{(1)}_{\rm min}$} and there are entangled regions in Fig.~\ref{fig_2}(b), (c), (d),(e) and (f). 

Let us now consider first Fig.~\ref{fig_1} and compare the detected entangled regions of the prethermalised and the genuine thermalised states. Here most of the entangled regions are overlapping, but at the surfaces there are extra regions of the genuine thermalized states. These regions are quite considerable for $\gamma$ close to $1$, see the results in Fig.~\ref{fig_1} (a) close to $h_0=0$. But even in this case there is an extra prethermalized region close to the critical point, $h_0=1$. We note that the detected entangled region shows non analytical behaviour, both for the prethermalized and the genuine thermalized states. By reducing the value of $\gamma$ the extra thermalized region shrinks, and in the case $\gamma \ll 1$ there is even an extra entangled prethermalized region detected by the negativity-based witness with {$\mu^{(2)}_{\rm min}$, see in Fig.~\ref{fig_1} (f). If we look at Fig.~\ref{fig_2}, the trend is rather the opposite. For larger values of $\gamma$ there are extra regions belonging to the prethermalized state. These extra regions start to shrink by decreasing $\gamma$ and for very small values of $\gamma$ the thermalized states have larger entangled regions. It is interesting to compare the entangled regions in the limit $\gamma \to 0^+$. In this limit in the prethermalized state just a part of the area is detected entangled, while in the thermalized state the complete area is entangled. 
\begin{figure*}[t]
\centering
  \begin{tabular}{@{}cc@{}}
\includegraphics[width=1.\columnwidth,angle=0]{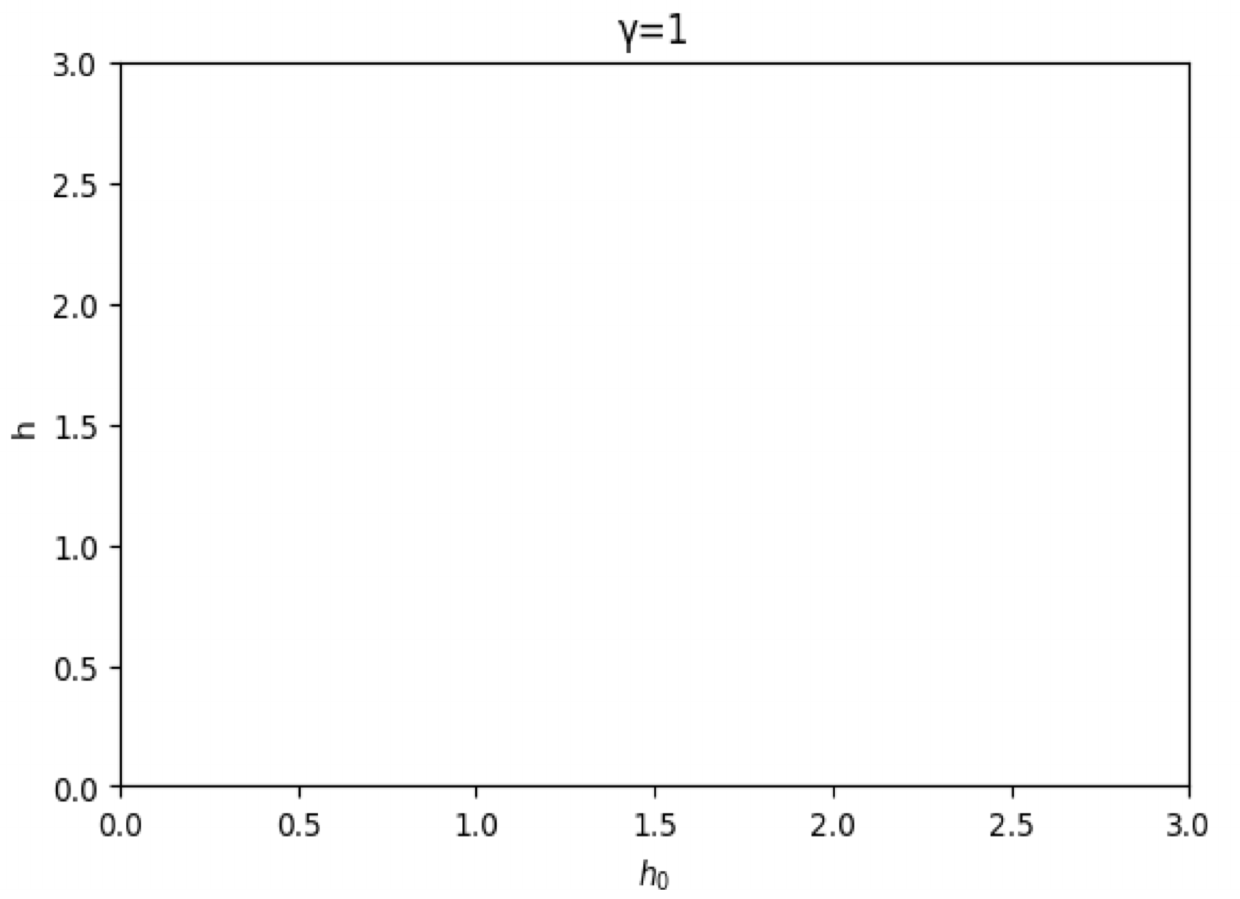} &
\includegraphics[width=1.\columnwidth,angle=0]{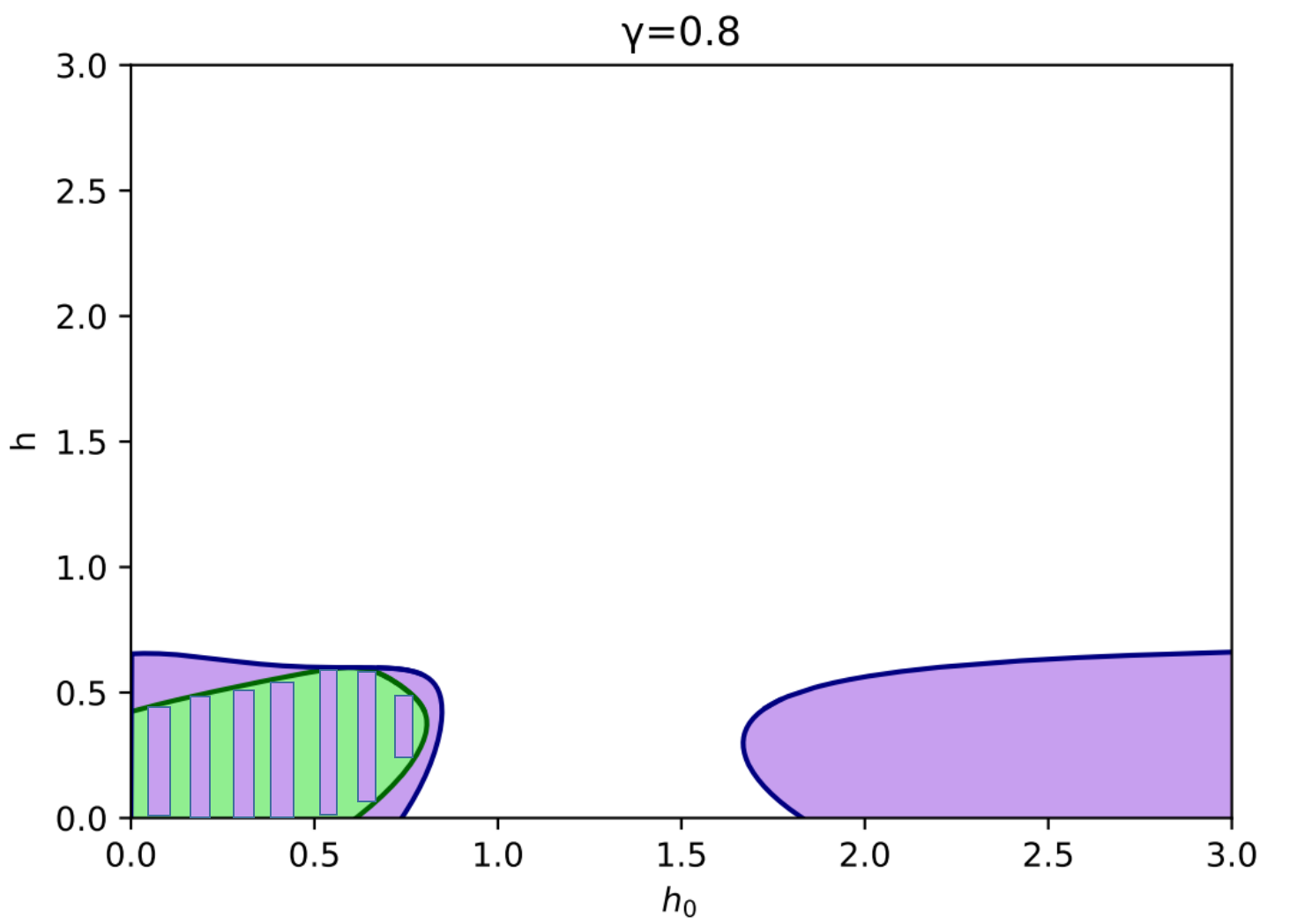} \\
\hspace{0.7cm}(a) & \hspace{1cm}(b)\\
&\\
\includegraphics[width=1.\columnwidth,angle=0]{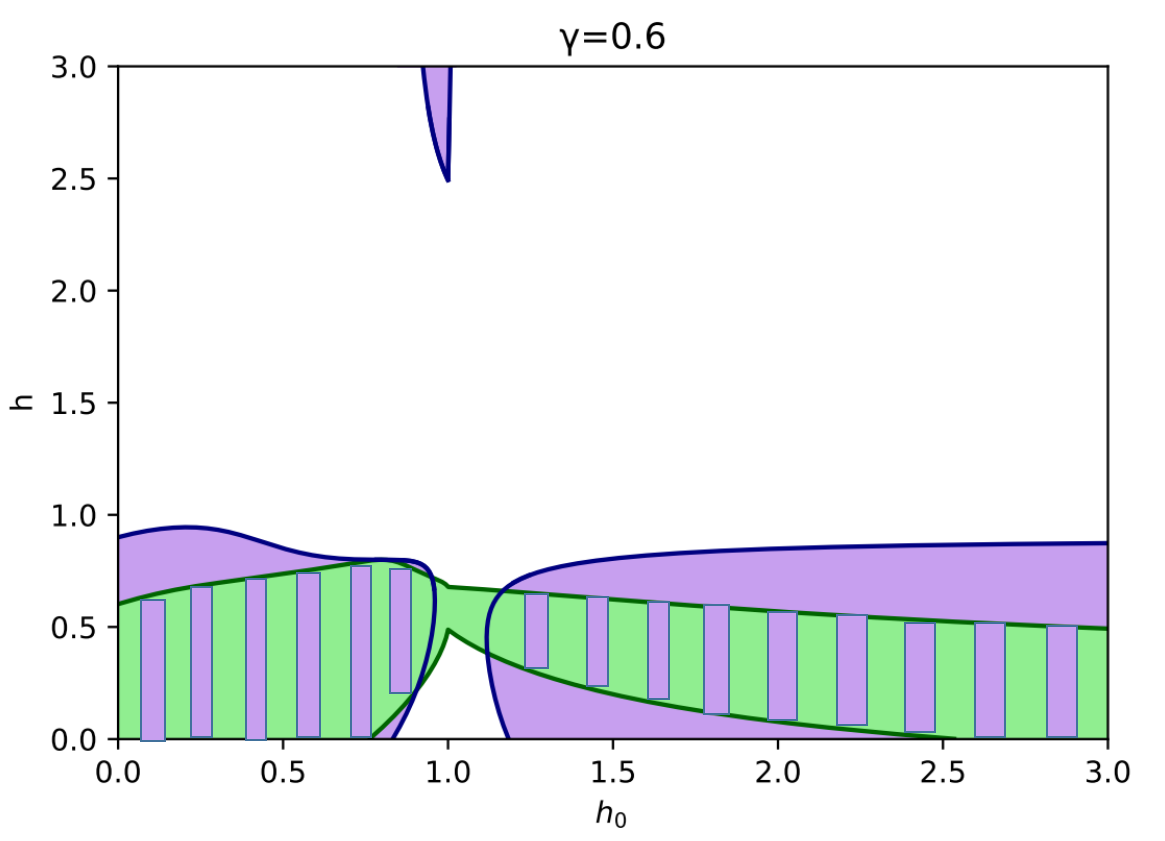} &
\includegraphics[width=1.\columnwidth,angle=0]{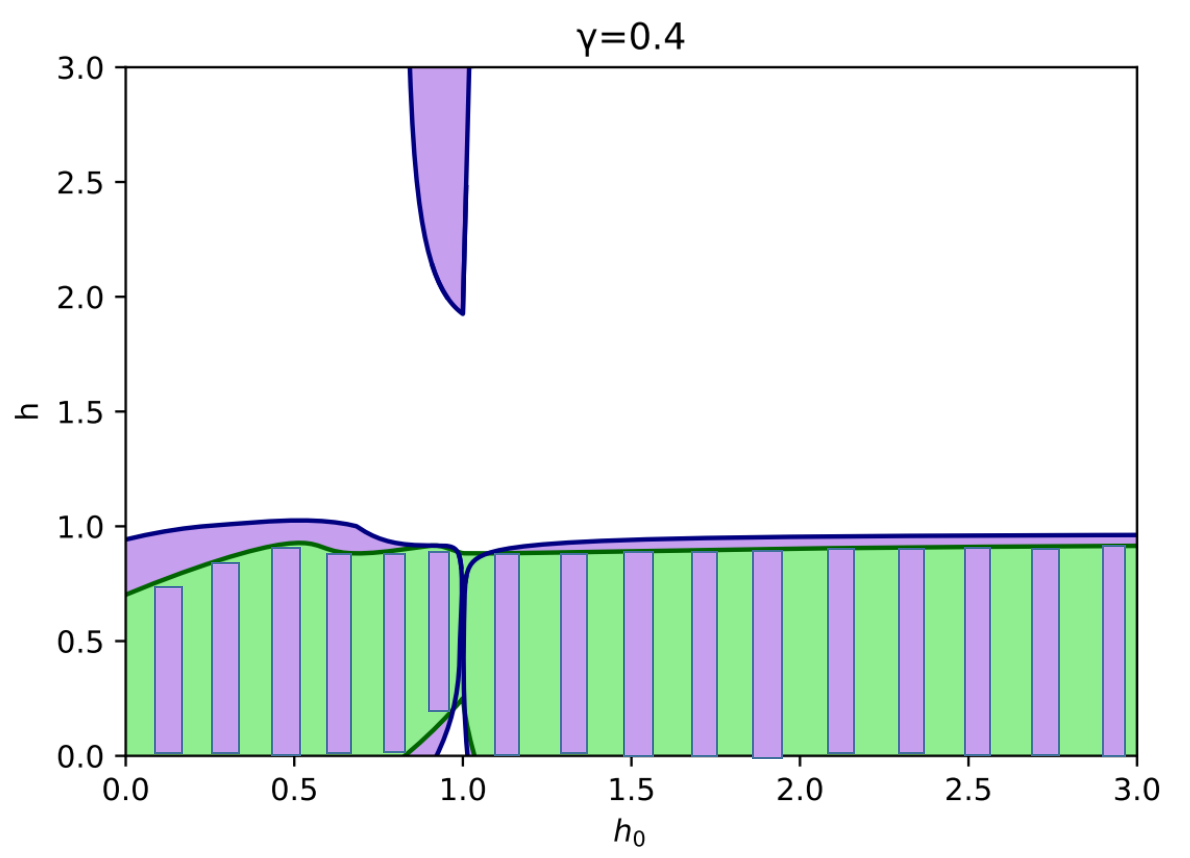} \\
\hspace{0.7cm}(c) & \hspace{1cm}(d)\\
&\\
\includegraphics[width=1.\columnwidth,angle=0]{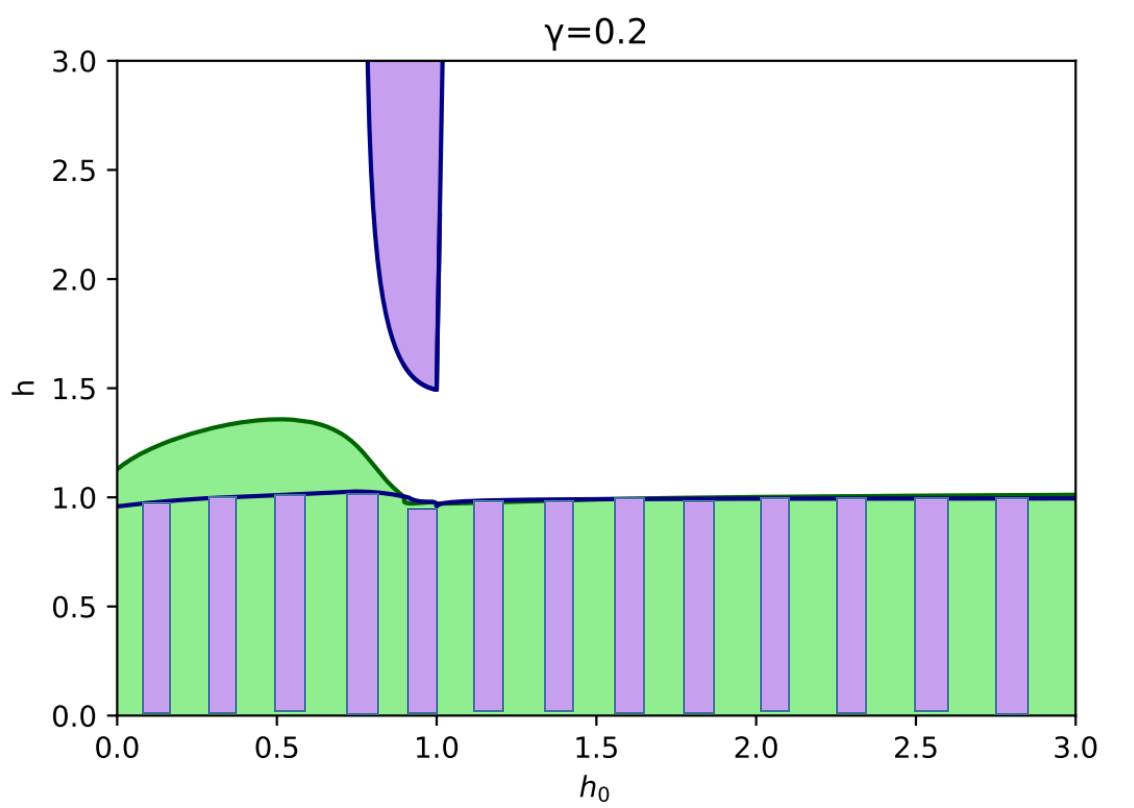} &
\includegraphics[width=1.\columnwidth,angle=0]{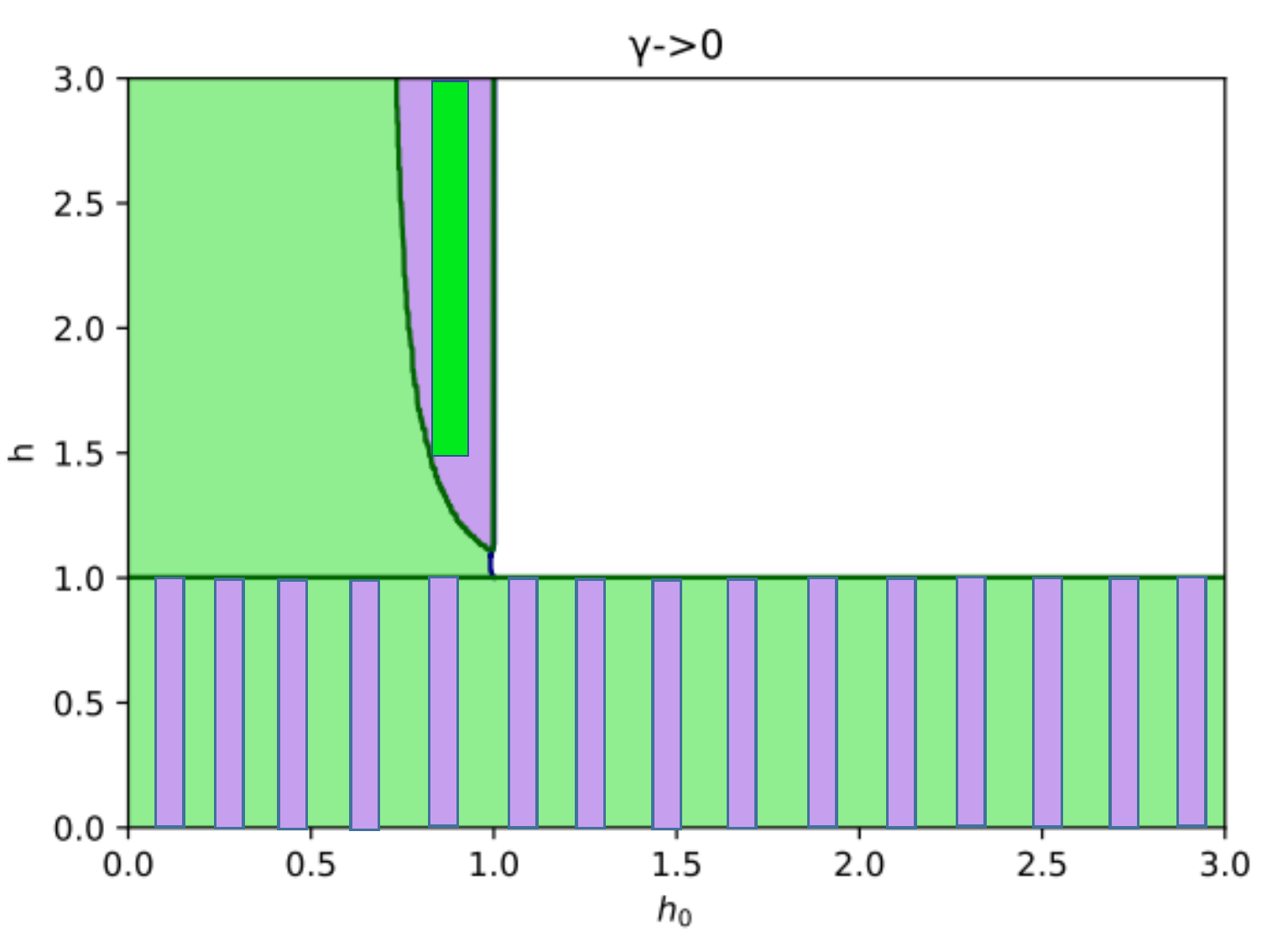} \\
\hspace{0.7cm}(e) & \hspace{1cm}(f)
  \end{tabular}
\caption{The same as in Fig.~\ref{fig_1} but using the negativity-based method with $\mu^{(2)}_{\rm min}$ in Eq.~(\ref{witness}).
\label{fig_2}}	
\end{figure*} 

\section{Discussion}
\label{sec:discussion}

Entanglement in mixed quantum states is a difficult problem, in particular when the degrees of freedom is large and we approach the thermodynamic limit. The possible systems of investigation are usually quantum spin chains, which could have experimental realizations in condensed matter systems  \cite{dutta_aeppli_chakrabarti_divakaran_rosenbaum_sen_2015} or they could be engineered artificially through ultracold atomic gases in an optical lattice. Recently, this latter type of technique is very well developed and different intriguing questions could be studied experimentally \cite{Greiner2002,Paredes2004,Sadler2006,PhysRevLett.98.160404,Kinoshita2006,Hofferberth2007,RevModPhys.80.885,Trotzky2012,Cheneau2012}. 

{In this paper we consider the $XY$ chain, which is integrable through free-fermionic techniques and several exact results are available, mainly in the ground state but there are some known results even at finite temperature \cite{PhysRevA.2.1075,PhysRevA.3.2137}. We consider the entanglement properties of mixed states of the $XY$ chain, which are non-equilibrium stationary states after a quantum quench protocol. To detect entanglement we use a family of entanglement witnesses that detect states with a nonzero bipartite entanglement negativity. In practice this witness  can detect all states that have nearest-neighbor entanglement.}

The mixed states we consider are due to a quench, when parameters of the Hamiltonian of the system are changed abruptly and the time evolution of the system is governed by the new Hamiltonian. After a sufficiently long time the system will approach a nonequilibrium stationary state, which is a mixed quantum state. For integrable systems, such as the $XY$ chain, the postquench state is described by a so-called Generalized Gibbs Ensemble, while for general, non-integrable systems it is expected to be a thermal state, which is described by an appropriate Gibbs ensemble. In expeiments one can not realize such systems, which are purely integrable, since weak integrable breaking perturbations are always present. If this perturbation is small the relaxation takes part in two steps. The system initially relaxes to a stationary state of the unperturbed Hamiltonian, which is the prethermalized state, while for later times genuine thermalization takes part. In the present paper we studied the entanglement properties of the two non-equilibrium stationary states, in particular we want to clarify the difference between the areas detected to be entangled by the entanglement negativity witness.

The entanglement negativity in Eq.(\ref{negativity}) for the $XY$-chain can be non-zero due to one of the two eigenvalues of the partial transpose, which are defined in Eqs.(\ref{eq:mumin1_corr}) and (\ref{witness}).  Therefore the areas which are detected entangled are indicated separately in Figs.~\ref{fig_1} and ~\ref{fig_2} for the two cases. We observed, that the areas corresponding to the prethermalized and the the genuine thermalized states are mainly overlapping, however there are extra regions at the boundaries of the overlapping areas. For the negativity witness in Eq.(\ref{eq:mumin1_corr}), corresponding to Fig.~\ref{fig_1}  generally the areas to be detected entangled increase during the thermalization process. Even in this case there are opposite tendencies close to the critical point of the initial state. Considering the other witness in Eq.(\ref{witness}) and the corresponding Fig.~\ref{fig_2}  here in the prethermalized state are larger entangled areas for larger values of $\gamma$, which trend however reverse for small values of $\gamma$. We mention that it would be interesting to check the entanglement properties of postquench states of other (Bethe-Ansatz) integrable models.

While we studied the nearest-neighbor entanglement of the postquench state, other properties uncovering hidden criticality of the initial system not detectable by local quantities have recently been considered \cite{Paul2022Hidden}. The method has been based on efficient lower bounds on the negativity in $XY$ chains \cite{Eisler_2015,PhysRevB.97.165123}. We have shown that the criticality of the initial state can still be seen in the boundaries of the regions with nearest-neighbor entanglement.

\begin{acknowledgments}
We thank G. T\'oth for previous cooperation in the project and for discussions. This work was supported by the Hungarian Scientific Research Fund under Grants No. K128989 and No. K146736 and by the National Research, Development and Innovation Office of Hungary (NKFIH) within the Quantum Information National Laboratory of Hungary.  \end{acknowledgments}

\bibliography{XYcomp}

\end{document}